\documentclass[10pt,journal,compsoc]{IEEEtran}
\usepackage{graphicx}
\usepackage{subfigure}
\usepackage{amsmath,amssymb}
\usepackage{booktabs}
\usepackage{cite}
\usepackage[T1]{fontenc} 
\ifCLASSINFOpdf
\else
\fi
\begin{document}
%
\title{Two-stream Network for ECG Signal Classification}
%
%
%

\author{Xinyao Hou, Shengmei Qin, Jianbo Su,~\IEEEmembership{Senior Member,~IEEE}

\IEEEcompsocitemizethanks{\IEEEcompsocthanksitem X. Hou and J. Su are with the Key Laboratory of System Control and
	Information Processing, Department of Automation, Ministry of Education,
	Shanghai Jiao Tong University, Shanghai 200240, China (e-mail: hxy0928@sjtu.edu.cn; jbsu@sjtu.edu.cn).\protect\\
	\IEEEcompsocthanksitem Qin is with Zhong Shan Hospital, No 180, Fenglin Road, Shanghai; 200032, China (e-mail: Qin.shengmei@zs-hospital.sh.cn).}
}

\maketitle

\begin{abstract}
Electrocardiogram (ECG), a technique for medical monitoring of cardiac activity, is an important method for identifying cardiovascular disease. However, analyzing the increasing quantity of ECG data consumes a lot of medical resources. This paper explores an effective algorithm for automatic classifications of multi-classes of heartbeat types based on ECG. Most neural network based methods target the individual heartbeats, ignoring the secrets embedded in the temporal sequence. And the ECG signal has temporal variation and unique individual characteristics, which means that the same type of ECG signal varies among patients under different physical conditions. A two-stream architecture is used in this paper and presents an enhanced version of ECG recognition based on this. The architecture achieves classification of holistic ECG signal and individual heartbeat and incorporates identified and temporal stream networks. Identified networks are used to extract features of individual heartbeats, while temporal networks aim to extract temporal correlations between heartbeats. Results on the MIT-BIH Arrhythmia Database demonstrate that the proposed algorithm performs an accuracy of 99.38\%. In addition, the proposed algorithm reaches an 88.07\% positive accuracy on massive data in real life, showing that the proposed algorithm can efficiently categorize different classes of heartbeat with high diagnostic performance.
\end{abstract}


%
\IEEEpeerreviewmaketitle

\section{Introduction}
%
%
%
%

\IEEEPARstart{T}{he} cardiovascular disease (CVD) is a major threat to human health and the leading cause of death internationally. According to the World Health Organization, ischaemic heart disease and all forms of stroke are responsible for a quarter of the total number of deaths globally \cite{meaning}. Electrocardiogram (ECG) records electrical activity generated by the heart. ECG is used for emotion recognition \cite{ECG} and biometric identification \cite{biometric}  due to its unique individual characteristics. ECG analysis is also an effective tool for assessing cardiac health, as it is easy to use as a non-invasive method. And ECG is one of the most effective way to study the classification of cardiac arrhythmias, which is of great interest to biomedical researchers. Therefore, accurate and effective analysis of ECG signals is critical in the diagnosis and treatment of cardiovascular disease \cite{New_problem}. However, there is a wide variety of heart diseases and a sheer volume of ECG data, which means that long-term reliance on a manual diagnosis not only easily introduces errors and leads to unstable diagnosis results, but also consumes a lot of medical resources. Therefore, there is an urgent need to realize a quick, accurate, and robust analysis of ECG data. Due to the time variation and unique individual characteristics of the ECG, the feature of the same kind of ECG signal varies in different patients under different physical conditions, which poses considerable difficulties for recognition of ECG patterns \cite{difficulty}. Many works have been carried out to advance the automatic analysis of ECG signals. In recent years, with the maturity and popularity of artificial intelligence technology, machine learning has become widely used in ECG signal analysis. The typical method is to divide the task into extraction and classification of ECG signal features. 

\begin{figure}[htbp]
	\centering
	\includegraphics[width=3.4in]{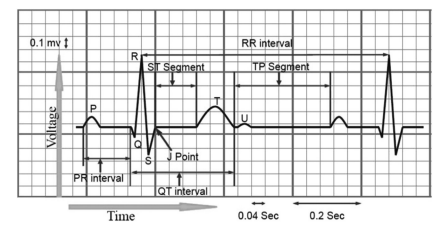}
	\caption{Standard fiducial points in the ECG. The ECG signal consists of five major deflections, including P, Q, R, S and T, plus a small deflection, known as the U wave \cite{survey}.}
	\label{ECG_signal}
\end{figure}

The correct representation of the ECG signal plays an important role in the diagnosis of heart disease, and different types of feature extraction techniques are used. In general, the P-QRS-T complex is the primary unit of the ECG signal \cite{complex}, as shown in Fig.~\ref{ECG_signal}. As the attributes of the P-QRS-T complex of healthy people are relatively fixed, it is often used as a reference by doctors when diagnosing the ECG signal, by comparing the normal values in the sinus rhythm for a healthy male adult with those of the patients \cite{PAMI1990}. Based on this approach, some researchers adopt a similar approach by using computer methods instead of manually extracting features of the QRS complex. There are many feature detection algorithms for the QRS complex, such as derivative-based \cite{derivative}, wavelet variation \cite{sahoo2020machine} and digital filters \cite{digital}. In addition, high-order moments are utlized to extract features of QRS complexes in \cite{high_oreder}. Some studies make use of statistical attributes of ECG signals, which provide the complexity and distribution of the ECG to achieve improved recognition performance \cite{statistical}. Frequency-domain-based techniques are also popular for classifying ECG signals \cite{WT}. And the ECG classification is achieved by various machine learning algorithms, represented by K-nearest neighbor (KNN) \cite{sahoo2020machine}, artificial neural network (ANN) \cite{ANN}, linear discriminant analysis (LDA) \cite{LDA} and support vector machine (SVM) \cite{SVM1}. In \cite{SVM_use}, a filtered feature selection method and SVM are combined to achieve early detection of ventricular fibrillation (VF) and rapid ventricular tachycardia (VT) in the MIT-BIH dataset. A block-based neural network (BbNN) is proposed in \cite{BbNN}, which is created by a group of bi-dimensional block networks with flexible structure and internal configuration. This work overcomes the possibility of varying ECG signals over time and individual differences, achieving 97\% classification accuracy on MIT-BIH.

Although these methods work well, they are difficult to implement and use because most are based on manually designed features. Different feature extraction techniques are used to cope with different diseases due to the importance of proper representation of the ECG signal. But there are many diseases and it is impossible to design features for each of them. Compared with traditional methods, deep learning can automatically extract features and perform complex data preprocessing. In recent years, deep learning has made impressive achievements in computer-aided medical diagnosis \cite{mine} and is used in ECG classification \cite{PAMI2021}. In \cite{2DCNN}, ECG signals are transformed into grayscale images and computer vision techniques are used. An end-to-end deep learning algorithm for ECG analysis is proposed, which employs a deep neural network (DNN) to classify  the class of ECG \cite{DNN}. In practice, many diseases need to be diagnosed on the holistic ECG signal, as some of them appear over time.  And most of the previous works focus on the classification of individual heartbeats instead of the holistic ECG signal.

There is very little effort devoted to classify the holistic ECG signal. Hence it motivates us to achieve the classification of the holistic ECG signal. As the ECG signal has temporal variation and unique individual characteristics, which means that the same type of ECG signal varies among patients under different physical conditions, a two-stream architecture is proposed in this paper. The architecture incorporates identified and temporal networks and accurately classify the holistic ECG over a long period. Specifically, individualized networks are used to extract features of individual heartbeats while temporal networks are employed to extract temporal correlations between heartbeats, taking into account temporal variation and unique individual characteristics of the ECG. And with the purpose of demonstrating the generalization and excellence of our architecture, seven detailed categories of heartbeat are collected, each containing data from a thousand different adults, also expending the ECG study with deep learning. The work operates as an inter-patient paradigm rather than an intra-patient paradigm. 

Section \ref{one} mainly introduces the proposed architecture for ECG recognition and provides a detailed description of its internal structure. In Section \ref{two}, the datasets uesd and the preprocessing are illustrated and the implementation details of the experiments are outlined. In Section \ref{three}, the performance of the architecture is evaluated in the MIT-BIH dataset and real life. The architecture is discussed in Section \ref{four} and this paper is summarized in Section \ref{five}.

\section{Two-stream Architecture for ECG Recognition} \label{one}

\begin{figure*}[!h]
	\centering
	\includegraphics[width=\textwidth]{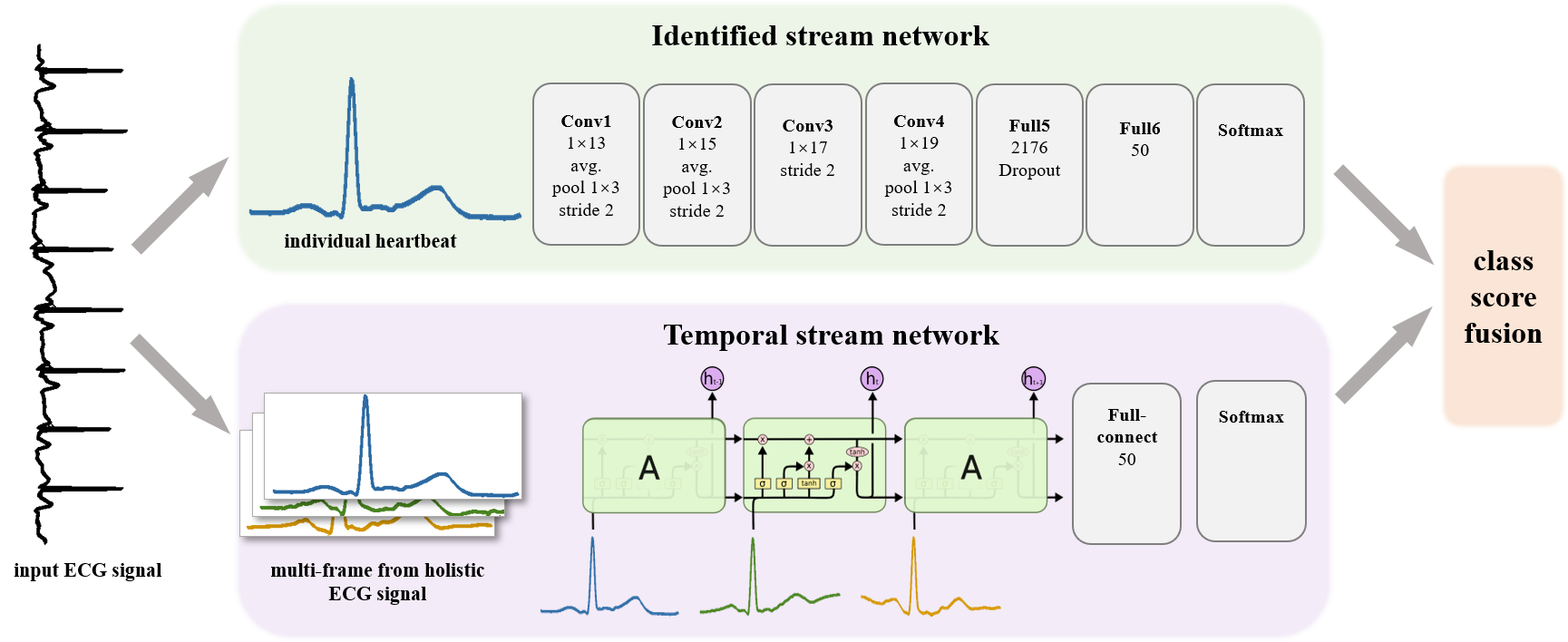}
	\caption{Two-stream architecture for ECG recognition}
	\label{frame}
\end{figure*}

The ECG signal is influenced by the object and time, which means that the same type of ECG signal varies among patients under different physical conditions. And inspired by the Two-Stream networks of action recognition\cite{twostream}, where the action is composed of spatial and temporal stream, the two-stream architecture of ECG recognition is proposed. In this architecture, the ECG signal can be decomposed into identified and temporal components. The identified part, in the form of an individual heartbeat (represented as the P-QRS-T complex), contains information about the unique individual characteristics represented in the ECG signal. The temporal part, in the form of a holistic ECG signal (represented as the combination of multiple P-QRS-T complexes), transmits the symptoms of the ECG and its changes over time. The corresponding ECG recognition architecture accordingly is designed and divided it into two streams, as shown in Fig.~\ref{frame}. Each stream is implemented with a neural network, whose result is combined by late fusion. Two fusion methods are considered: averaging the output scores and training a fully connected layer on the stacked features extracted from each stream.

Most studies in the field of heartbeat classification focus on individual heartbeats and use an intra-patient paradigm. In the scheme, the heartbeats of the same patient used in both training and testing subsets make the evaluation result overly optimistic \cite{optmistic}. Moreover, this scheme does not take into account the time variation and unique individual characteristics of the ECG. Our architecture takes these aspects into consideration and operates as an inter-patient paradigm rather than an intra-patient paradigm. 

\subsection{Identified stream network}

The ECG signal contains unique individual characteristics and disease symptoms. The identified stream network operates at an individual heartbeat, efficiently extracting the identity features and static features of the heartbeat. The appearance of the individual heartbeat (static characteristics of the heartbeat) is a useful clue as many diseases can be identified from a single heartbeat without the holistic ECG signal. The classification of individual heartbeats can be achieved by the identified stream network designed. And the classification of individual heartbeats is pretty competitive for some specific classes. On the other hand, the symptoms vary among patients, identity features matter. This network can be pre-trained for the purpose of identification in order to realize the identification capability on its own.

\begin{figure}[htbp]
	\centering
	\includegraphics[width=3.4in]{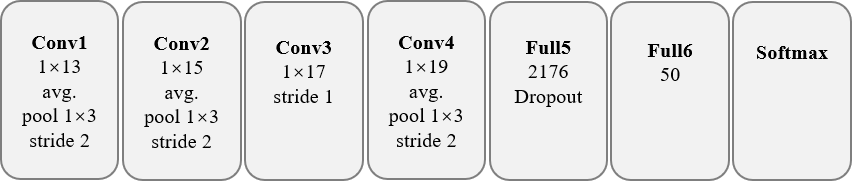}
	\caption{The architecture of proposed identified stream network}
	\label{identified}
\end{figure}

The identified stream network is constructed by an 11-layer 1D convolution neural network \cite{1DNN}. It consists of seven alternating convolutions and average-pooling layers, followed by a dropout layer and two fully-connected layers, as shown in Fig.~\ref{identified}. The 1D convolution layers are used to extract features of the heartbeat in different receptive fields, the output expression is shown as follows, 
\begin{equation} 
	h_{i}^{l, k}=f\left(b_{i}^{l, k}+\sum_{n=0}^{N} W_{}^{l, k} * h_{i+n}^{l-1, k}\right)\label{1Dnn},
\end{equation}
where $h_{i}^{l, k}$ is the $i^{th}$ output of layer $l$, $f()$ is the activation function, $b_{i}^{l, k}$ is the offest of the $i^{th}$ neuron in layer $l$, $W_{}^{l, k}$ is the $k^{th}$ convolution kernals in layer $l$.
The average-pooling layers preserve the overall features of the input signal and enhance the robustness of the network. The dropout layer and fully-connected layers are used to generalize features flexibly and reduce overfitting.

\subsection{Temporal stream network}
The temporal stream network exploits the information of temporal variations in the heartbeat. As mentioned above, the P-QRS-T complex is the primary unit of the ECG signal. It means that a P-QRS-T complex can be regarded as an individual heartbeat while the holistic ECG signal can be regarded as the sequential combination of multiple complexes in the time dimension. As a video is composed of multiple frames, each individual heartbeat can be mapped to a frame, and the holistic ECG signal can be regared as a video sequence in the lower dimension. In this way, the overall features of the target signal can be explored for the temporal correlation between heartbeats with the method in the video comprehension field. And the temporal correlation between heartbeats is significant. For example, sinus tachycardia is usually determined by  the variations in the interval between multiple heartbeats and the temporal correlation between heartbeats, instead of the features of a single heartbeat. In this paper, long short-term memory (LSTM) \cite{LSTM} is used to extract the temporal correlation between heartbeats, and the output expression is shown in \eqref{LSTM}. And the output is considered a feature and used for the heartbeat classification with the fully-connected layer and Softmax.
\begin{equation} \label{LSTM}
	\left\{
	\begin{aligned}
		f_{t} &=\sigma\left(W_{f} \cdot\left[h_{t-1}, x_{t}\right]+b_{f}\right), \\
		i_{t} &=\sigma\left(W_{i} \cdot\left[h_{t-1}, x_{t}\right]+b_{i}\right), \\
		\tilde{C}_{t} &=\tanh \left(W_{C} \cdot\left[h_{t-1}, x_{t}\right]+b_{C}\right), \\
		C_{t} &=f_{t} * C_{t-1}+i_{t} * \tilde{C}_{t}, \\
		h_{t} &=\sigma\left(W_{o}\left[h_{t-1}, x_{t}\right]+b_{o}\right) * \tanh \left(C_{t}\right), \\
	\end{aligned}
	\right.
\end{equation}
where $C_{t}$ is the output of cell state in moment $t$, $h_{t}$ is the output of hidden state in moment $t$. $x_{t}$ is the input in moment $t$, representing the waveform of the $t$th heartbeat ('frame') of the whole input ECG signal. The specific selection method of $x_{t}$ is introduced in the next subsection.

\subsection{Selecting `frame' of ECG signal} \label{select}
The ECG signal is strictly continuous, unlike video where the distribution of frames is time-discrete. Two methods to select `frame' in the ECG signal are proposed. One is to view each heartbeat beat as a frame, that is, the frame is selected by the P-QRS-T complex. The center of the frame is R peak, which is composed of 0.25 seconds before R peak and 0.35 seconds after R peak. As the sampling rate is 500 Hz, there are 300 time-points of amplitude in one frame. In this method, the intervals between heartbeats are ignored as most of this part remains unchanged and is lack of information for classification. If this method is applied, the temporal stream (LSTM) is used to detect the minor changes in the trajectory of consecutive frames, ignoring the potential changes in intervals. The other method is to split the ECG signal chronologically, treating each segment as a frame and there is the same time interval between each segment. In this method, there are also 300 time-points of amplitude in each frame as the former method to reduce the influence of unchanged part in intervals. If this method is applied, the temporal stream (LSTM) is used to detect the change of intervals between each P-QRS-T complex as the coordinates of R peak in each frame changes with the heartbeat interval change. However, when extreme cases occur, there is no guarantee that each frame will contain a single heartbeat.

\begin{figure*}[htbp]
	\centering
	\includegraphics[width=\textwidth]{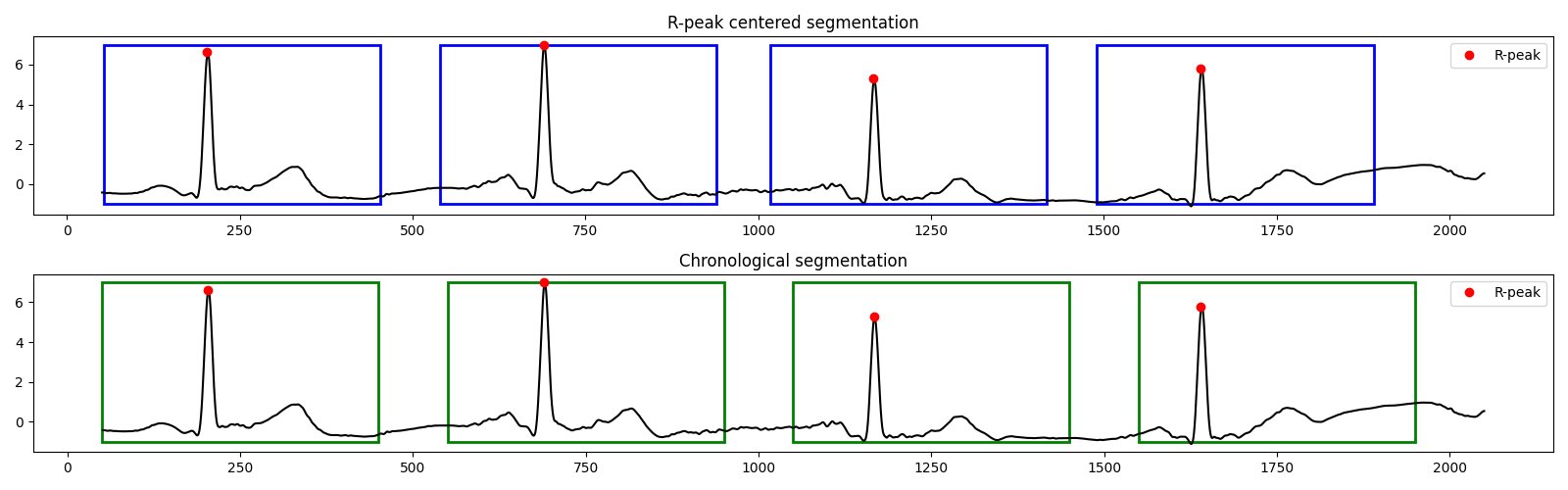}
	\caption{Two methods of selecting frame from ECG signal}
	\label{segmentation}
\end{figure*}

In most cases, the results of these two methods are largely consistent, as normal heartbeat intervals are regular. However, for some diseases with variations in heart rate, the choice of method can have an impact on the results. Results from these two methods are compared in Section \ref{Evaluation}. The two methods are shown in Fig.~\ref{segmentation}.

\section{Experiments} \label{two}

\subsection{Dataset}
The Massachusetts Institute of Technology-Beth Israel Hospital (MIT-BIH) Arrhythmia database is widely used in research for analysis of ECG signal \cite{dataset}. The dataset includes 48 two-lead ECG records obtained from 47 persons, with a duration of 30 minutes for each of them. The sampling rate is 360 Hz with the resolution for digitization of 11-bit over the range of 10-mV. The MIT-BIH contains a wide range of annotated information that can be used to optimize relevant algorithms. Most prior efforts are carried out based on MIT-BIH but are limited by the small number of patients and rhythm types present in it. There are 109449 individual heartbeats included which are grouped into five different classes: normal (N), supraventricular ectopic beat (S), ventricular ectopic beat (V), fusion beat (F) and unknown beat (Q).

A novel ECG dataset with a large number of patients is constructed in this study, which is annotated by experts for a broad range of ECG rhythm classes. The dataset is designed to achieve the classification of 7 rhythms from raw 8-lead ECG inputs comprised of 7000 ECG records from 7000 patients based on it, each of them having a duration of 10 seconds. The sampling rate is 500 Hz, and all the data is de-identified. The names of 7 classes mentioned are sinus tachycardia (ST), conduction block (CB), complete right bundle branch block (RBBB), left ventricular hypertrophy (LVH), frequent premature ventricular contractions (PVC), right axis deviation (RAD) and health control (HC). All the datasets are divided into train-sets, test-sets and valid-sets in the ratio of 7:2:1.

\subsection{Pre-process}

The ECG signals are collected in a complex environment mixed with various interference, which brings different types of noise and artifacts to the ECG records. Typical noise can be grouped as power line interference, baseline wander, electrosurgical noise and electromyography noise\cite{noise}. The bandpass filter, low-pass filter and wavelet transform are usually used to delete artifact signals and achieve the denoising of ECG signals. The wavelet transform (WT)-based down-sampling and resampling are used to delete artifact signals from ECG signals in this paper. The adaptive threshold filtering algorithm is used in this filter and function ‘db8’ is selected as the wavelet function. This paper only carries out simple denoising methods to reduce data distortion while enhancing the generalization of the network. 

\begin{figure}[htbp]
	\centering
	\includegraphics[width=3.4in]{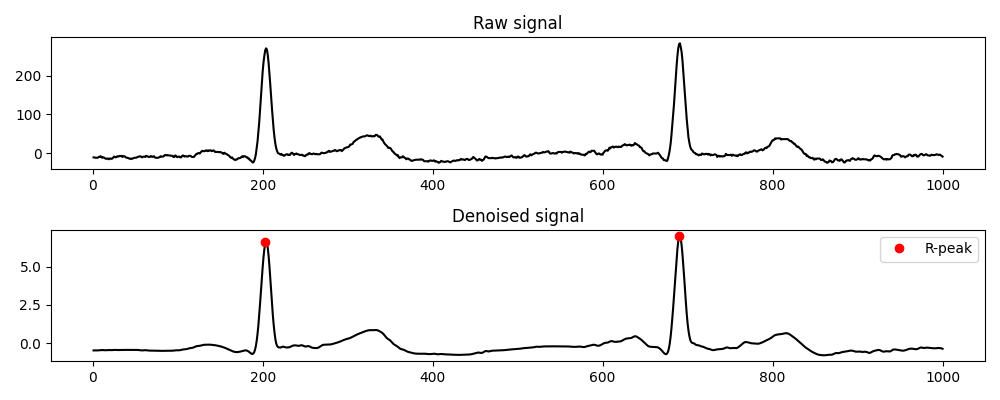}
	\caption{The raw ECG signal and signal after denoising and normalizing}
	\label{denoised}
\end{figure}

The P-QRS-T complex is the main component unit of ECG signal and is used for the segmentation of individual heart-beat of ECG signals. In the MIT-BIH, each signal is marked with an annotation including the disease classification and the location of the R-peak of each individual heartbeat. In our dataset, the improved Pan-Tompkins algorithm is used to detect R-peak \cite{Rpeak}. The heartbeat segments can be determined by the R-peaks for further use. All the segments are normalized with the Z-score method. Fig.~\ref{denoised} shows the ECG signal before and after denoising.

\subsection{Implementation details}
In this paper, experiments are carried out on MIT-BIH Arrhythmia Database and our independent dataset. The sampling rate of the two datasets is different. A segmentation containing 300 sampling points is selected as an individual heartbeat with R-peak as the center, including the whole P-QRS-T complex, and achieves consistency in the input dimensions of the two datasets.

Identified stream network is implemented on both MIT-BIH and our dataset for the classification of the individual heartbeat. There are 76614 individual heartbeats grouped into five classes detected from train-sets of MIT-BIH and 54616 individual heartbeats grouped into seven classes detected from train-sets of our database. Each individual heartbeat is fed to this network with the dimension of 1*300. The average-pooling is performed over 1*3 windows with stride 2. All hidden weight layers use the rectification activation function. The learning rate decay, learn rate drop period, and learning rate parameters are set to 0.1, 20, and 0.01 within a total of 60 epochs with a batch size of 1000. And Adam optimizer is used. There are various works carried out for achieving heartbeat classification in MIT-BIH, and the result comparison is listed in Section \ref{Evaluation}.

The temporal stream network is operated in our dataset for the classification of holistic ECG signals. As there are 7000 single-lead ECG signals in our dataset but only 48 in MIT-BIH, 10 frames are selected from each ECG signal as an input signal of the network with two different methods mentioned above. If there are less than ten frames that can be selected, such as not detecting ten individual heartbeats, the input signal is complemented with zeros. The learning rate decay, learn rate drop period, and learning rate parameters are set to 0.1, 20, 0.01 in a total of 60 epochs with a batch size of 600. And the Adam optimizer is used.

The whole architecture is implemented in our dataset for the better classification of holistic ECG signals. Identified stream network is pre-trained on MIT-BIH for extracting unique individual characteristics, as each ECG record has a duration of 30 minutes and can provide enough individual heartbeats for training. The temporal stream network is pre-trained in our dataset same as described in the previous paragraph. A joint stack of a fully-connected layer on top of the output of the penultimate fully-connected layer of each stream is trained to realize the classification. The input of identified stream network in the architecture is the first frame of each ECG signal and the input of the other stream is 10 frames from each ECG signal. We also try another late fusion by averaging the output scores of the first frame and the holistic 10 frames, and in this fusion individual characteristics are ignored.

\section{Evaluation of Proposed Model}	\label{three}
\subsection{Evaluation index}
In this paper, confusion matrix, accuracy ($Acc$), sensitivity ($Sen$), precision ($Ppv$) and $F1$ score are used to measure model accuracy, using the cardiologist committee annotations as the ground truth. The corresponding expressions are as follows,
\begin{equation} \label{index}
	\left\{
	\begin{aligned}
		A c c &=\frac{T P+T N}{T P+T N+F P+F N}, \\
		S e n &=\frac{T P}{T P+F N}, \\
		P p v &=\frac{T P}{T P+F P}, \\
		F1 &= \frac{Ppv * Sen * 2}{Ppv + Sen},
	\end{aligned}
	\right.
\end{equation}
where $TP$, $TN$, $FN$ and $FP$ stands for true positive, true negative, false positive and false positive, respectively.

\subsection{Evaluation results of proposed model}\label{Evaluation}
First, the performance of the identified stream network is measured in the MIT-BIH dataset to prove the capacity of our model. The task is classifying five heartbeat classes: N, S, V, F and Q. \cite{kalfilter,BbNN,martis2012,martis2013,1DNN2,2DCNN} are carried out for achieving this task, and their results are listed in Table \ref{mitresult1} with the comparison to ours. The accuracy rate of our network is 99.38\%, performing as well as the other methods. Table \ref{mitresult2} illustrates the specific result of our model predictions on MIT-BIH. The precision ($Ppv$) of fusion beat (F) is only 91.57\% and the mistakes are not surprising because the number of Fusion beats is smaller than other classes. The precision ($Ppv$) and recall ($Seb$) of the other heartbeat classes reach 99\%, which shows that our network has a good performance both on the overall MIT-BIH and on specific heartbeat classes. The result also proves that deep learning has the ability of ECG classification.

\begin{table*}[h]
	\caption{The performance comparison with other algorithms of heartbeat classification on MIT-BIH}
	\label{mitresult1}
	\begin{center}
		\setlength{\tabcolsep}{15mm}{
			\begin{tabular}{ccc}
				\hline
				{Author}         &Method 					& {$Acc$} \\ \hline
				Sayadi et al.\cite{kalfilter}     &Kalman filter              & 99.10\%           \\
				Shadmand et al.\cite{BbNN}     &BbNN              & 98\%           \\
				Martis et al.\cite{martis2012}       &PCA          	& 98.11\%        \\
				Martis et al.\cite{martis2013}       &PCA + PNN classifier         	& 99.58\%        \\
				Wu et al.\cite{1DNN2}         &Wavelet+CNN          & 97.2\%  \\
				Jun et al.\cite{2DCNN}        &picture-based + 2D-CNN         & 99.05\%            \\
				\textbf{Our model}            &\textbf{11-layer 1D-CNN}			& \textbf{99.38\%}     \\
				\hline
		\end{tabular}}
	\end{center}
\end{table*}

\begin{table}[h]
	\caption{The overall classification performance of our model on MIT-BIH}
	\label{mitresult2}
	\begin{center}
		\setlength{\tabcolsep}{5mm}{
			\begin{tabular}{cccc}
				\hline
				& {$Ppv$} 	& {$Sen$}	& {$F1-score$}		\\ \hline
				N    & 99.67\%    & 99.59\%      & 99.63\%\\
				S    & 99.94\%    & 99.63\%  	& 99.78\%\\
				V    & 99.86\%    & 99.44\%  	& 99.65\%	\\
				F    & 91.57\%    & 92.48\%  	& 92.02\%        \\
				Q    & 98.07\%    & 98.75\%  	& 98.41\%				\\
				\hline
				average    & 99.38\%    & 99.34\%  	& 99.36\% \\
				\hline\hline
				overall $Acc$                            &&&99.38\% \\
				\hline
		\end{tabular}}
	\end{center}
\end{table}

Then, the identified stream network is also performed in real life to show the generalization of the model. Our dataset is constructed from real data with more micro heartbeat classes which is most common in real life. These bring great challenge to classification of ECG signals. The task is to classify seven micro-class of heartbeat with 1000 patients in each, including sinus tachycardia (ST), conduction block (CB), complete right bundle branch block (RBBB), left ventricular hypertrophy (LVH), frequent premature ventricular contractions (PVC), right axis deviation (RAD) and health control (HC). Table \ref{ourresult1} lists the specific results. The accuracy rate of our network on our dataset is 78.29\% and all other evaluation indicators are also reduced compared to Table \ref{mitresult2}. This is because our dataset does not contain the prior information on the R-peak position, , which is also unknown in advance in real life, and the detection of the R-peak position by the algorithm will introduce errors. Moreover, the differences among heartbeat classes in our dataset are smaller than those of MIT-BIH, but the number of classes is larger. These result in the higher difficulty of achieving ECG classification in our dataset. The difficulty of ECG classification on our dataset is also confirmed by transforming other methods into our dataset. As a result, further work is carried out on our dataset to reveal the improvement.

\begin{table}[h]
	\caption{The overall performance of identified stream on our dataset}
	\label{ourresult1}
	\begin{center}
		\setlength{\tabcolsep}{5mm}{
			\begin{tabular}{cccc}
				\hline
				& {$Ppv$} 	& {$Sen$}	& {$F1-score$}		\\ \hline
				ST    & 79.97\%    & 90.83\%      & 85.05\%\\
				CB    & 79.59\%    & 79.00\%  	& 79.29\%\\
				RBBB    & 84.74\%    & 74.21\%  	& 79.13\%	\\
				LVH    & 89.21\%    & 65.25\%  	& 75.37\%        \\
				PVC    & 75.03\%    & 76.35\%  	& 75.69\%				\\
				RAD    & 89.06\%    & 80.06\%  	& 84.32\%				\\
				HC    & 58.76\%    & 77.36\%  	& 66.79\%				\\
				\hline
				average    & 79.48\%    & 77.58\%  	& 77.95\% \\
				\hline
				\hline
				overall $Acc$                            &&&78.29\% \\ \hline
		\end{tabular}}
	\end{center}
\end{table}

\begin{figure}[!h]
	\centering
	\subfigure[]{
		\begin{minipage}[t]{0.5\linewidth}
			\centering
			\includegraphics[width=1.8in]{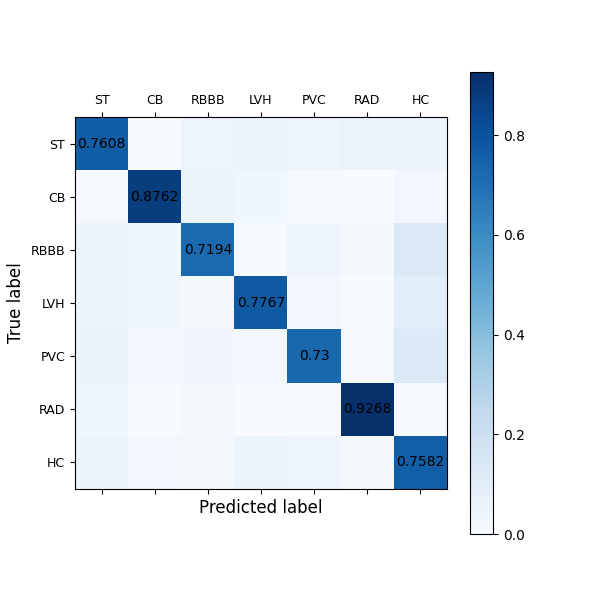}
		\end{minipage}%
	}%
	\subfigure[]{
		\begin{minipage}[t]{0.5\linewidth}
			\centering
			\includegraphics[width=1.8in]{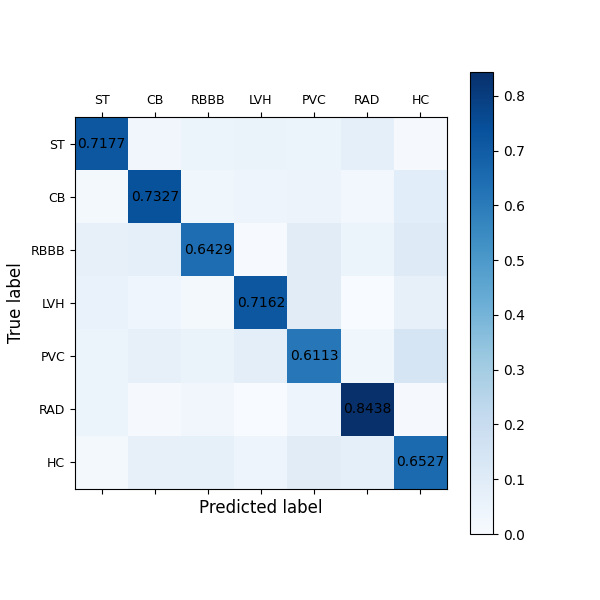}
		\end{minipage}%
	}%
	\centering
	\caption{ Confusion Matrix of  temporal stream network. (a) is the result of R-peak centered 'frame' selecting method and (b) is the result of chronological segmentation method. The accuracy of each class is displayed on a color gradient scale. }
	\label{CM1}
\end{figure}

The temporal stream network is evaluated which is designed to classify the holistic ECG signal. The effectiveness of the input configurations described in Section \ref{select} is firstly assessed. Table \ref{ourresult2} and Fig.~\ref{CM1} compare the performance of two methods. The results illustrate that the R-peak centered method has a better performance in this task, whose classification accuracy is 79.36\% and is far better than chronological segmentation method. This is because the chronological segmentation method may involve an incomplete heartbeat, which only matters when meet detecting heart rate variability-related diseases and is not applicable in most cases. As a result, R-peak-centered segmentation is chosen in late fusion, the overall accuracy of which achieves 79.36\% and the specific results are listed in Table \ref{ourresult3}. The result also proves that the proposed temporal stream network has the ability to classify the holistic ECG signal correctly, and its performance needs to be improved.

\begin{table}[!h]
	\caption{Result comparison of selecting frame methods on our dataset}
	\label{ourresult2}
	\begin{center}
		\setlength{\tabcolsep}{10mm}{
			\begin{tabular}{cc}
				\hline
				{Method}          & {$Acc$} \\ \hline
				R-peak centered                   & 79.36\%           \\
				Chronological                 	& 69.64\%        \\
				\hline
		\end{tabular}}
	\end{center}
\end{table}

\begin{table}[!h]
	\caption{The overall performance of temporal stream on our dataset}
	\label{ourresult3}
	\begin{center}
		\setlength{\tabcolsep}{5mm}{
			\begin{tabular}{cccc}
				\hline
				& {$Ppv$} 	& {$Sen$}	& {$F1-score$}		\\ \hline
				ST    & 73.27\%    & 76.08\%      & 74.65\%\\
				CB    & 87.19\%    & 87.62\%  	& 87.41\%\\
				RBBB    & 78.77\%    & 71.94\%  	& 75.20\%	\\
				LVH    & 82.90\%    & 76.77\%  	& 80.20\%        \\
				PVC    & 81.56\%    & 73.00\%  	& 77.04\%				\\
				RAD    & 90.05\%    & 92.68\%  	& 91.35\%				\\
				HC    & 63.30\%    & 75.82\%  	& 69.00\%				\\
				\hline
				average    & 79.58\%    & 79.26\%  	& 79.26\% \\
				\hline
				\hline
				overall $Acc$                            &&&79.36\% \\ \hline
		\end{tabular}}
	\end{center}
\end{table}

\begin{table}[!h]
	\caption{Result comparison of different selecting frame methods}
	\label{finalresult1}
	\begin{center}
		\setlength{\tabcolsep}{10mm}{
			\begin{tabular}{cc}
				\hline
				{Method}          & {$Acc$} \\ \hline
				fusion by averaging                   & 80.21\%           \\
				fusion by fc layer                 	& 88.07\%        \\
				\hline
		\end{tabular}}
	\end{center}
\end{table}

\begin{table}[!h]
	\caption{The overall performance of two-architecture on our dataset}
	\label{finalresult2}
	\begin{center}
		\setlength{\tabcolsep}{5mm}{
			\begin{tabular}{cccc}
				\hline
				& {$Ppv$} 	& {$Sen$}	& {$F1-score$}		\\ \hline
				ST    & 83.68\%    & 87.88\%      & 85.73\%\\
				CB    & 91.51\%    & 93.80\%  	& 92.64\%\\
				RBBB    & 90.41\%    & 83.08\%  	& 86.59\%	\\
				LVH    & 88.88\%    & 87.10\%  	& 87.98\%        \\
				PVC    & 91.81\%    & 86.30\%  	& 88.97\%				\\
				RAD    & 95.36\%    & 96.60\%  	& 95.98\%				\\
				HC    & 76.14\%    & 81.70\%  	& 78.82\%				\\
				\hline
				average    & 88.26\%    & 88.07\%  	& 88.10\% \\
				\hline\hline
				overall $Acc$                            &&&88.07\% \\ \hline
		\end{tabular}}
	\end{center}
\end{table}

\begin{figure}[!h]
	\centering
	\includegraphics[width=2.5in]{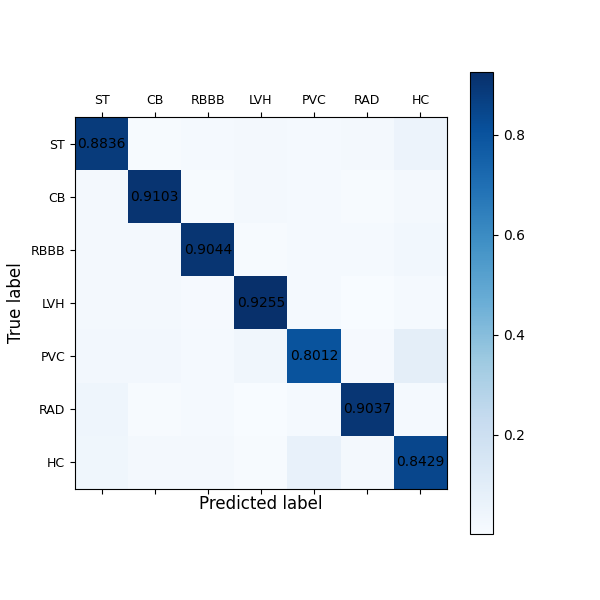}
	\caption{Confusion Matrix of our two-stream architecture, the accuracy of each class is displayed on a color gradient scale.}
	\label{finalpic}
\end{figure}

Finally, the complete two-stream model is evaluated, which combines the two streams to improve the performance of ECG classification. Table \ref{finalresult1} presents the result comparison of two late fusion methods. Re-training a fully connected layer on stacked features from two streams performs better, with 7.86 percentage points higher. We are also surprised by the improvement. In this method, identified stream network is pre-trained on the dataset for the purpose of identification, which illustrates that the individual characteristics matter in ECG signal classification. The final result of our two-stream architecture is shown in Table \ref{finalresult2} and its confusion matrix is shown as Fig.~\ref{finalpic}. These results convince us that our architecture performs well in ECG signal classification and is competitive, and there is a large room for improvement with our architecture.

\section{Discussion} \label{four}
To sum up, we have developed and validated a new architecture to achieve the classification of holistic ECG signals in this study. The two-stream architecture is widely used in action recognition, where the action is considered to be determined by the instantaneous spatial distribution and temporal variation. Thus, the two-stream architecture in action recognition is composed of the spatial and temporal streams and performs well. The ECG signal is determined by patients and varies over time. Because of the similarity between ECG classification and action recognition, the Two-Stream Architecture is used to address both the influences accordingly. The architecture incorporates identified and temporal stream networks, extracting individual heartbeat features and temporal correlations between heartbeats respectively. The fusion of features enables the classification of ECG signals. 

The classification of cardiac arrhythmias by ECG is of great interest for biomedical researchers. The relevant studies have been carried out for a long time. And thanks to the efforts of researchers, the accuracy of classification is extremely high when meeting the five macro heartbeat classes. The identified stream in our architecture also performs well as shown in Table \ref{mitresult1}. However, it is not sufficient to classify only five macro heartbeat classes in real life and we pursue the classification of micro heartbeat classes. The result of the classification of six main micro heartbeat classes with our identified stream is shown in Table \ref{ourresult1}, and all the evaluation indicators are reduced compared to Table \ref{mitresult2}. The precision of HC is only 58.76\%, which means that many patients are misdiagnosed to be healthy and the results are not satisfactory. There is no doubt that the other methods mentioned in Table \ref{mitresult1} will also not perform well on the set of data collected in real life. Indeed, all these methods only take into account the classification of individual heartbeats and ignore the global consideration of the holistic ECG signal. When meeting the task of classifying micro heartbeat classes, the individual heartbeat features are far from sufficient and the temporal variations in heartbeats also matter. Therefore, classification of the ECG signal holistically rather than an individual heartbeat is considered an appropriate choice. On the other hand, the classification of individual heartbeats always uses an intra-patient paradigm while the classification of holistic ECG signal is operated as an inter-patient paradigm, which is more convincing. 

The regular method to classify the holistic ECG signal with deep learning is to design a very deep neural network, like \cite{DNN}. But the length of ECG signal is set using this method because the dimension of input is constant. And the depth of the neural network should increase with the length of ECG signal, which means that this method cannot achieve the classification of the long ECG signals. Thus, the typical recurrent neural network, LSTM, is chosen as the temporal stream to classify the holistic ECG signal. As the result shown in Table \ref{ourresult3}, the temporal stream can classify micro heartbeat classes. The result is not satisfactory as the stream is designed mainly to extract temporal correlations between heartbeats. And it will have a better performance when more tricks are utilized. Before the late fusion of two streams, the identified stream is pre-trained with the purpose of identification on the MIT-BIH database. Because the global structure remains, the stream also retains the ability to classify individual heartbeats, and the ability to extract identity features is given through pre-training. As the result shown in Table \ref{finalresult2}, the accuracy is improved when the two streams are fused. It illustrates that the architecture of the two-stream has a good effect on improving ECG classification by comprehensively considering the two influencing factors of ECG signals. And there is no doubt that our architecture has a large room for improvement, as there are several limitations in our model. First is the network chosen in the temporal stream, although LSTM is designed for sequences, it is not suitable when the period of ECG signals is too large (such as 24h). Second, our method is based on single-lead ECG recordings, and it is necessary to introduce a method for multiple-lead ECG recordings. 

All in all, our model is an end-to-end model for holistic ECG classification that can analyze patients' ECG signals directly from raw data. The heartbeat classes in our dataset are micro-classes of arrhythmias that are difficult to distinguish but important and rarely studied. The method can be used for real-time ECG monitoring and early warning on light portable devices. Last but not least, it is proved that some doctors can deduce the potential threats of the other organs through the ECG signals, and our final purpose of studying ECG signals is to determine the organs of disease through heartbeats. It can improve health surveillance capability and reduce the use of medical resources. The work in this paper has revealed the ability to accurately categorize heart diseases using computer analysis, and to conclude the organs of disease with ECG signals is feasible. Further efforts will be made to identify the diseased organs through heartbeats.

\section{Conclusions} \label{five}
A deep ECG signal classification model with a two-stream architecture is proposed in this paper, which is based on the time variation and unique individual characteristics of the ECG. The model incorporates identified and temporal recognition streams based on neural networks. The identified stream network extracts the identity features and static features of the heartbeat, while temporal stream network monitors the temporal variations of the ECG signals. The proposed model takes various influencing factors of ECG into consideration, and effectively realizes the classification of the individual heartbeat and the holistic ECG signal. Experiments are carried out on the MIT-BIH Arrhythmia Database and our independent dataset, it appears that the identified stream in our proposed model performs as well as similar work. And when the two streams are combined, the proposed model provides higher diagnostic performance, achieving an accuracy rate of 99.38\% on MIT-BIH and 88.07\% on our independent dataset. A great deal still needs to be done in the future. The computer-aided diagnosis of ECG effectively realizes heart health monitoring and promotes the development of intelligent medical treatment.

\ifCLASSOPTIONcaptionsoff
  \newpage
\fi



%
\bibliographystyle{IEEEtran}
\bibliography{./reference}

\end{document}